\documentclass[fleqn,usenatbib]{mnras}

\usepackage{newtxtext,newtxmath}
\usepackage[T1]{fontenc}
\usepackage{ae,aecompl}
\usepackage{graphicx}
\usepackage{amsmath}
\usepackage{booktabs}
\usepackage{multirow}
\usepackage{xcolor}
\usepackage{hyperref}
\usepackage{orcidlink} 

\hypersetup{
	colorlinks=true,
	citecolor=blue,
	urlcolor=blue,
	linkcolor=blue
}

\title[CatBoost vs SED-Fitting for Galaxy Properties]{CatBoost versus  Spectral Energy Distribution‑Fitting: Estimating Galaxy Properties under Controlled Photometric Incompleteness}

\author[V. Asadi, H. Haghi and AH. Zonoozi]{
	Vahid Asadi\,\orcidlink{0009-0005-8897-2385}$^{1}$\thanks{E-mail: vahijd.asadij@gmail.com},
	Hosein Haghi\,\orcidlink{0000-0002-9058-9677}$^{1,2,3}$ and
	Akram Hasani Zonoozi\,\orcidlink{0000-0002-0322-9957}$^{1,2}$
	\\
	$^{1}$Department of Physics, Institute for Advanced Studies in Basic Sciences (IASBS), PO Box 11365-9161, Zanjan, Iran\\
	$^{2}$Helmholtz-Institut f\"ur Strahlen-und Kernphysik (HISKP), Universit\"at Bonn, Nussallee 14-16, D-53115 Bonn, Germany\\
	$^{3}$School of Astronomy, Institute for Research in Fundamental Sciences (IPM), PO Box 19395 - 5531, Tehran, Iran
}

\date{Accepted XXX. Received YYY; in original form ZZZ}

\pubyear{2026}

\begin{document}
	\label{firstpage}
	\pagerange{\pageref{firstpage}--\pageref{lastpage}}
	\maketitle
	

\begin{abstract}
Estimating galaxy physical parameters from photometric data is fundamentally challenged by missing measurements that are endemic to astronomical surveys. Using a mock catalog from the Horizon‑AGN hydrodynamical simulation that provides the true physical parameters, we evaluate CatBoost, a gradient‑boosting algorithm that natively handles missing data, under a deliberately adversarial scenario: we train it on 12 photometric bands with increasing levels of injected missingness (10\%, 20\%, 30\% missing per band) and compare its performance against an idealised parametric spectral energy distribution (SED)‑fitting reference that uses complete 26‑band photometry (no missing data, all bands available). This asymmetric design represents an upper‑bound, best‑case baseline for traditional methods. Despite this intentional disadvantage, CatBoost’s performance degradation is limited: moving from complete data to 30\% missing, mass RMSE increases from 0.08 to 0.18 dex, SFR RMSE from 0.41 to 0.53 dex, and redshift RMSE from 0.20 to 0.28, while bias remains near zero. Against the ideal SED‑fitting reference, CatBoost trained on only 12 bands with 30\% missing values achieves lower errors for mass (0.18 vs. 0.28 dex) and SFR (0.53 vs. 0.57 dex) and removes the systematic biases present in the SED‑fitting results. For redshift, the SED‑fitting reference has lower NMAD (0.030 vs. 0.093 at extreme missingness) while CatBoost maintains smaller bias. These results suggest that CatBoost’s native handling of missing values can offer practical advantages for extracting galaxy properties from imperfect photometric surveys, at least under the conditions explored here.
\end{abstract}



\section{Introduction}
Understanding the physical properties of galaxies across cosmic time is a central objective of modern extragalactic astronomy. Among these properties, stellar mass, star formation rate (SFR), and redshift are key parameters used to trace galaxy growth and star formation across cosmic time \citep[e.g.,][]{brinchmann2004physical, noeske2007star, bolzonella2010tracking, madau2014cosmic}. The extraction of these properties from observational data hinges almost entirely on spectral energy distribution (SED)-fitting to a range of stellar population models \citep[e.g., ][]{Bruzual2003,maraston2005evolutionary,conroy2009propagation}.

While physically interpretable, SED-fitting is fundamentally limited by model assumptions, degeneracies between parameters, and its sensitivity to photometric uncertainties \citep[e.g., ][]{conroy2009propagation,Acquaviva2015,Pacifici2015,carnall2018inferring}. These challenges become increasingly severe as surveys expand to millions of sources, where the computational cost of running full SED fits becomes challenging \citep[e.g.,][]{Hemmati2019,asadi2025leveraging}. Compounding this issue, missing photometric data is inherently non-random and correlated with galaxy properties: band-dependent depths, filter coverage, and wavelength-dependent SED shapes produce systematic patterns of non-detections. Because these missing values correlate with intrinsic galaxy properties and observational conditions, they introduce non-random gaps in the available information and further destabilize SED-based inference.

In recent years, machine learning (ML) techniques have emerged as powerful alternatives for galaxy property estimation, offering speed, non-parametric flexibility, and often superior accuracy and bias reduction \citep[e.g.,][]{Hemmati2019,simet2021comparison,Davidzon2022,chartab2023machine,la2024estimating,asadi2025leveraging,Asadi_2025,AsadiETG_2026}. A common approach is to train ML algorithms on synthetic catalogs and then apply them to real data. For instance, \cite{Asadi_2025} used CatBoost \citep{prokhorenkova2018catboost} to classify quiescent and star‑forming galaxies, finding that the ML sample was more complete than those derived from parametric SED‑fitting or colour‑colour techniques. Importantly, while that work employed a specialized imputation method to fill missing values, CatBoost itself natively handles missing data by treating them as a distinct feature category during tree construction.

This study provides a deliberately asymmetric stress test of CatBoost against a parametric SED‑fitting method for galaxy property estimation (mass, SFR and redshift) under controlled missingness. Using a mock catalog from the Horizon‑AGN simulation that contains the true physical parameters for each galaxy \citep{dubois2014dancing}, we inject missing values with a magnitude‑dependent sigmoid function, calibrating each band to achieve three controlled missingness levels (10\%, 20\%, and 30\% of galaxies missing per band). We then train CatBoost on 12 photometric bands (a realistic subset) with these increasing missing fractions, and compare its performance to an idealised, upper‑bound reference: the parametric SED‑fitting estimates (\texttt{LePhare} \citep{Arnouts1999, Ilbert2006}) as provided in the Horizon‑AGN catalog, which were derived using all 26 available bands with no missing data. This asymmetric design--superior input data for the traditional method versus deliberately disadvantaged input for the ML method – is intentional. It allows us to answer a practically relevant question: can a modern ML algorithm with native missing‑data handling match or surpass a parametric template‑based method even when the ML method is given substantially less and sparser information? We also examine how CatBoost’s internal feature importance shifts as missingness increases, which helps explain its robustness, and we test the scalability of this behaviour from mild to extreme missingness.

The paper is organized as follows. Section \ref{sec:2} describes the Horizon-AGN data utilized in this work. Section \ref{sec:3} details the methodology for creating the progressive missing data catalogs, the CatBoost framework, and our analysis techniques. Section \ref{sec:4} presents the results, detailing the feature reweighting mechanism and the comparative performance against SED-fitting. Finally, we discuss the implications of our findings in Section \ref{sec:5} and conclude in Section \ref{sec:6}.

We adopt a flat $\Lambda$CDM cosmology with parameters $\text{H}_{0}=70\text{kms}^{-1}\text{Mpc}^{-1}$, $\Omega_{\text{m}}=0.3$, and $\Omega_{\Lambda}=0.7$. All magnitudes are reported in the AB system \citep{oke1983secondary}.


\section{Data}\label{sec:2}

This study uses a mock galaxy catalog derived from the Horizon-AGN hydrodynamical simulation \citep{dubois2014dancing}. The catalog, presented by \cite{laigle2019horizon}, contains 789,354 galaxies extracted from a $1 \times 1$ deg$^2$ lightcone using the AdaptaHOP halo finder \citep{aubert2004origin} applied to the stellar particle distribution. Each stellar particle (with mass $\sim 2 \times 10^6 \text{M}_\odot$) is associated with a synthetic simple stellar population using the \cite{Bruzual2003} (BC03) models, assuming a Chabrier initial mass function (IMF) \citep{Chap2003} and interpolating between the metallicity values available in BC03. 
The galaxy sample is selected based on stellar mass ($\text{M} > 10^9 \text{M}_\odot$) and redshift ($0 < \text{z} < 4$).

The Horizon-AGN virtual observatory reproduces the optical and near-infrared (NIR) photometry of COSMOS2015 galaxies \citep{Laigle2016}, including the ten broad bands (u, B, V, r, $\text{i}^{+}$, $\text{z}^{++}$, Y, J, H, $\text{K}_\text{s}$), the fourteen medium-band filters (Subaru/SuprimeCam; \citep{taniguchi2007cosmic}) and the two Spitzer/IRAC channels at 3.6 $\mu$m and 4.5 $\mu$m (denoted ch1 and ch2).

For each filter, the signal-to-noise ratio (SNR) distribution is matched to that of the ultra-deep stripes of COSMOS2015. Observational uncertainties are incorporated by perturbing the original galaxy fluxes according to their SNR. The model accounts for attenuation by both dust and the inter-galactic medium, though it does not include flux contamination from nebular emission.

The galaxy sample is selected from the Horizon-AGN mock catalog using the following criteria:

\begin{itemize}
    \item \textit{SNR limit}: SNR $>$ 1.5 in all photometric bands to exclude non-detections (e.g., when the observed flux is smaller than the flux error),
    \item \textit{NIR detection limit}: $\text{K}_\text{s} < 24.7$, matching the depth of the COSMOS2015 ultra-deep stripes.
    \item \textit{Photometric bands used}: the ten broad bands plus the two Spitzer/IRAC channels (total of 12 bands).
\end{itemize}

Applying these criteria yields a final sample of 519,758 galaxies with complete photometric measurements across all bands.


\begin{figure*}
	\centering
	\includegraphics[width=0.9\linewidth]{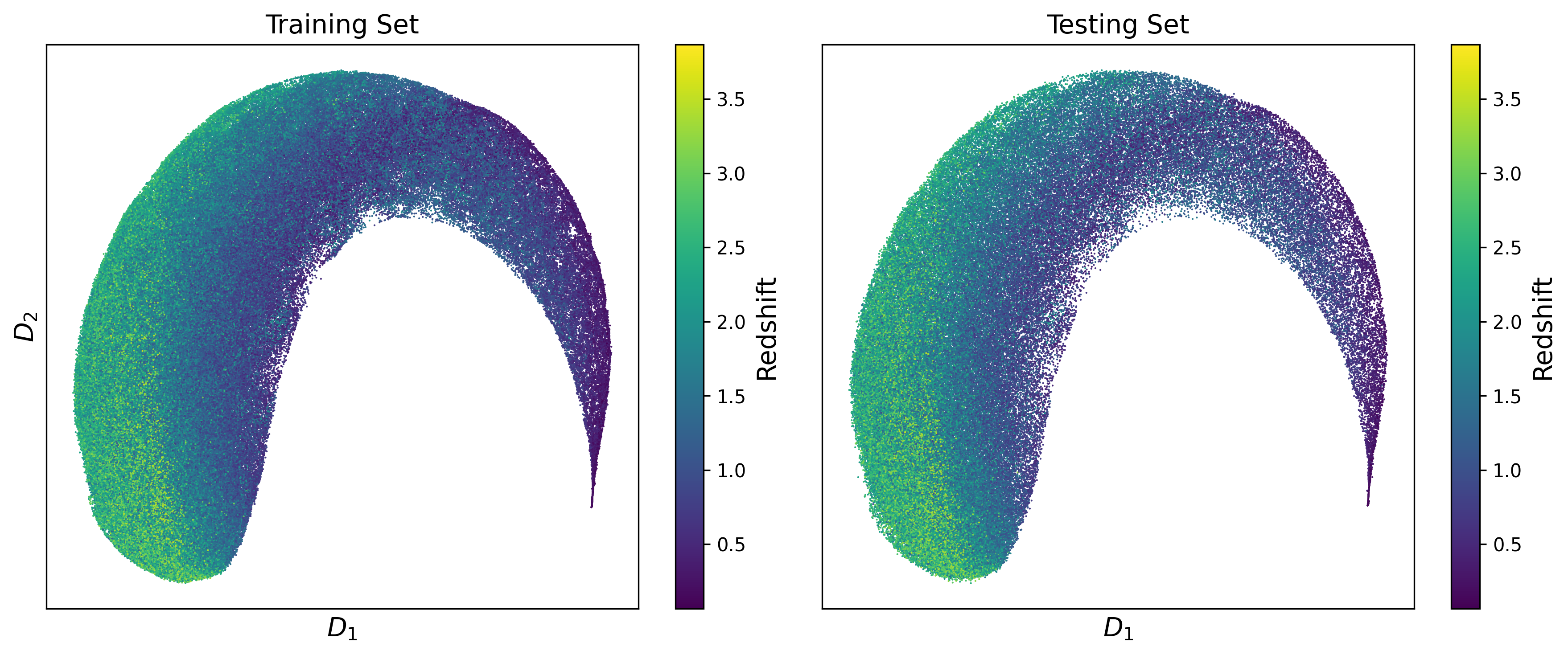}
	\caption{UMAP visualization of training (70\%) and testing (30\%) sets colored by redshift, demonstrating the representative split used for model development and evaluation. UMAP was fitted to the training set and the testing set was then projected into the same embedding using the trained UMAP transformation.}
	\label{fig:fig1}
\end{figure*}

\section{Methodology}\label{sec:3}
\subsection{Missing Data Transfer}\label{sec:3.1}
We generate incomplete photometric datasets by removing galaxy fluxes in a magnitude‑dependent manner, simulating the real detection limits of a survey. The probability that a galaxy’s magnitude \(\text{m}\) in a given band is missing follows a logistic (sigmoid) function:

\begin{equation}
	\text{P}_{\text{missing}}(\text{m}) = \frac{1}{1 + e^{-\alpha (\text{m} - \text{m}_{50})}},
\end{equation}

where we fix \(\alpha = 1.5\) (a typical transition width) and \(\text{m}_{50}\) is the magnitude at which the probability of being missing is 50\%. Bands with shallower depth have brighter \(\text{m}_{50}\), so fainter galaxies are more likely to be missing – a behavior directly analogous to real astronomical imaging surveys.

For each band we calibrate \(\text{m}_{50}\) to achieve a desired global missing fraction. To study the adaptation mechanism of CatBoost under increasing data loss, we define three levels:

\begin{itemize}
	\item \textit{Mild:} \(\approx 10\%\) of galaxies missing per band,
	\item \textit{Moderate:} \(\approx 20\%\) missing per band,
	\item \textit{Extreme:} \(\approx 30\%\) missing per band.
\end{itemize}

For each level, we determine the required \(\text{m}_{50}\) value for every band by matching the mean of \(\text{P}_{\text{missing}}(\text{m})\) over the galaxy sample to the target fraction. The injection is performed independently for each level, and all twelve bands are made incomplete simultaneously within a level. Missing values are represented as \texttt{NaN} in the sample. A fixed random seed (42) ensures reproducibility. We note that this injection is a controlled simplification of real survey incompleteness; the limitations of this approach are discussed in Section~\ref{sec:5.2}.

\subsection{Machine Learning Framework}\label{sec:3.2}
We employed the \texttt{CatBoostRegressor} algorithm \citep{prokhorenkova2018catboost} with the \texttt{MultiRMSE} loss function for our multivariate regression task, training a single ensemble of trees to predict mass, SFR, and redshift simultaneously. This multivariate approach allows the model to learn the inherent correlations among these physical parameters, potentially leading to more physically consistent predictions compared to training separate models for each parameter. 

CatBoost has demonstrated strong performance in astronomical applications, including galaxy classification \citep{Asadi_2025}, quasar identification \citep{hughes2022quasar}, and source characterization \citep{coronado2022classification}, among other studies \citep{cunha2022photometric, humphrey2023euclid, coronado2023redshift, zeraatgari2024exploring, boulet2024catalogue, li2025application}.

CatBoost is a gradient-boosting algorithm that builds an ensemble of decision trees sequentially, with each new tree trained to correct the errors of the previous trees. This iterative approach progressively minimizes a loss function, enhancing the model's predictive capability. A key advantage of CatBoost is its native handling of missing values, which aligns directly with our focus on incomplete photometric data. It treats missing values as a separate, informative state during tree construction. At each split in a tree, the algorithm learns the optimal direction—left or right child node—to send all observations with missing values in that feature, based on which path yields the best gain. This integrated method allows missingness itself to become a useful signal for prediction, eliminating the need for manual imputation.

We randomly divided the galaxy sample into training (70\%) and testing (30\%) sets, resulting in 363,830 and 155,928 galaxies respectively, using a fixed random seed to ensure reproducible splits across all experiments. Because the missing values were injected into the full dataset prior to splitting (Section~\ref{sec:3.1}), the training and testing sets each contain the same missingness pattern. Figure~\ref{fig:fig1} provides a visual comparison of the training and testing sets using Uniform Manifold Approximation and Projection \citep[UMAP; ][]{mcinnes2018umap}, colored by redshift, demonstrating that the random split preserved similar redshift distributions across training and testing data.

The model was trained exclusively on the training set, with performance evaluation conducted on the held-out testing set. For each of the four data conditions (complete, mild, moderate, extreme; see Section~\ref{sec:3.1}), we performed an independent hyperparameter optimisation using a randomised search with 5‑fold cross‑validation over 50 iterations \citep{pedregosa2011scikit}. The search targeted the same three key hyperparameters: number of iterations (100–2000), learning rate (0.01–0.3), and tree depth (3–10). The best configuration for each scenario is listed in Table~\ref{tab:hyperparameters}. The model was then retrained on the full training set of that scenario using the corresponding optimal hyperparameters, and evaluated on the testing set. Early stopping was not used; the number of iterations was determined by the randomised search.

For training we used the 12 photometric bands described in Section~\ref{sec:2} for their transmission curves). This filter set approximates a realistic, widely used combination in extragalactic surveys.

\begin{table}
	\centering
	\caption{Optimal hyperparameters found by randomised search for each data scenario.}
	\label{tab:hyperparameters}
	\begin{tabular}{l|ccc}
		\hline\hline
		Data Scenario & Iterations & Learning Rate & Tree Depth \\
		\hline
		Complete      & 1620 & 0.084 & 8 \\
		Mild (10\% missing)  & 1868 & 0.059 & 7 \\
		Moderate (20\% missing) & 1341 & 0.081 & 7 \\
		Extreme (30\% missing) & 1138 & 0.043 & 9 \\
		\hline
	\end{tabular}
\end{table}

\subsection{SED‑Fitting Reference}\label{sec:3.3}
We adopt the mass, SFR, and redshift estimates from the Horizon‑AGN catalog \citep{laigle2019horizon} as an idealised, best‑case reference for traditional parametric SED‑fitting. These values were derived by the catalog creators using the \texttt{LePhare} SED‑fitting code following the same methodology as the COSMOS2015 catalog. The fitting used the BC03 stellar population models with a Chabrier IMF and two parameterisations of SFHs (exponentially declining and delayed models). Critically, this reference was produced using all 26 photometric bands (the ten broad bands, fourteen medium bands, and two IRAC channels) with no missing data. We did not perform any additional SED‑fitting ourselves; the catalog values are used as provided. This represents an optimistic upper bound on what a traditional parametric method can achieve under perfect data conditions.

In deliberate contrast, our CatBoost model is trained on 12 broad bands (the 10 optical/NIR bands plus the two IRAC channels) and on data with artificially injected missingness (mild, moderate, extreme levels). Thus, the comparison is asymmetric by design: the SED‑fitting reference receives richer, complete information, while CatBoost receives poorer, incomplete information. We therefore compare CatBoost (disadvantaged) against this idealised reference to assess whether native missing‑value handling can compensate for information loss.

\subsection{Analysis Framework}\label{sec:3.4}
To quantitatively track how CatBoost adapts to missing data, we used the model's built-in feature importance \citep{prokhorenkova2018catboost}. We adopt the default CatBoost importance type, \texttt{PredictionValuesChange}, which measures the average change in the model's predictions when a feature's value is altered. In this approach, CatBoost accumulates the total reduction in the training loss attributable to splits on each feature across all trees in the ensemble and then normalizes these values so that the importances sum to unity. This yields a scalar importance score for each photometric band that reflects its overall contribution to improving the model’s predictions.

These feature importance scores provide direct interpretability by indicating how strongly the trained model relies on each band when constructing its decision trees. They enable comparative analysis by tracking importance shifts across different missingness scenarios and reveal how the model progressively redistributes reliance from increasingly incomplete bands toward more stable ones.

To evaluate predictive accuracy, we compare the predicted values of both CatBoost and SED‑fitting against the true physical parameters (mass, SFR, redshift) from the Horizon‑AGN simulation. We use the following metrics:

We report the Root Mean Square Error (RMSE), which captures the typical deviation of the predictions from the true values:

\begin{equation}
	\text{RMSE} = \sqrt{\frac{1}{\text{N}} \sum_{\text{i}=1}^{\text{N}} (\text{y}_\text{i} - \hat{\text{y}}_\text{i})^2},
\end{equation}

where $\text{y}_i$ is the true value, $\hat{\text{y}}_i$ is the predicted value, and N is the number of galaxies.

The bias, or mean error, indicates the systematic offset of the predictions:

\begin{equation}
	\text{Bias} = \frac{1}{\text{N}} \sum_{\text{i}=1}^{\text{N}} (\text{y}_\text{i} - \hat{\text{y}}_\text{i}).
\end{equation}

Lastly, the normalised median absolute deviation (NMAD) – a robust measure of scatter, less sensitive to outliers:

\begin{equation}
	\text{NMAD} = 1.48 \times \operatorname{median}\left(\frac{|\text{y}_{\text{i}} - \hat{\text{y}}_\text{i}|}{1+\text{y}_{\text{i}}}\right)
\end{equation}

These metrics are computed separately for CatBoost (under complete and incomplete data scenarios) and for the SED‑fitting reference, providing a comprehensive evaluation of both scatter and systematic bias across all missing data scenarios.


\section{Result}\label{sec:4}
We first compare the accuracy and bias of CatBoost (trained on 12 bands with increasing missingness) against the idealised parametric SED‑fitting reference (complete 26‑band photometry) for galaxy property estimation under controlled photometric completeness (0\% missing) and three levels of incompleteness: mild ($\approx$10\%), moderate ($\approx$20\%), and extreme ($\approx$30\% missing per band). We then examine how CatBoost’s internal feature importance shifts as missingness increases, which helps explain its robustness.

\begin{table}
	\centering
	\caption{Mass estimation performance across methods and data scenarios. The metrics are in dex.}
	\label{tab:mass_performance}
	\begin{tabular}{l|c|ccc}
		\hline\hline
		Method & Data Scenario & \text{RMSE} & Bias & \text{NMAD} \\
		\hline
		SED‑fitting & Complete & 0.275 & –0.226 & 0.029 \\
		\hline
		\multirow{4}{*}{CatBoost} & Complete (0\%)      & 0.079 &  -0.000 & 0.006 \\
		& Mild (10\%)   & 0.114 &  -0.000 & 0.008 \\
		& Moderate (20\%)& 0.147 &  0.000 & 0.009 \\
		& Extreme (30\%)& 0.178 &  0.000 & 0.010 \\
		\hline
	\end{tabular}
\end{table}

\begin{table}
	\centering
	\caption{SFR estimation performance across methods and data scenarios. The metrics are in dex.}
	\label{tab:sfr_performance}
	\begin{tabular}{l|c|ccc}
		\hline\hline
		Method & Data Scenario & \text{RMSE} & Bias & \text{NMAD} \\
		\hline
		SED‑fitting & Complete & 0.571 & 0.106 & 0.267 \\
		\hline
		\multirow{4}{*}{CatBoost} & Complete (0\%)     & 0.413 & -0.002 & 0.131 \\
		& Mild (10\%)   & 0.463 & -0.002 & 0.142 \\
		& Moderate (20\%)& 0.497 & -0.002 & 0.157 \\
		& Extreme (30\%)& 0.534 & -0.003 & 0.172 \\
		\hline
	\end{tabular}
\end{table}

\begin{table}
	\centering
	\caption{Redshift estimation performance across methods and data scenarios.}
	\label{tab:redshift_performance}
	\begin{tabular}{l|c|ccc}
		\hline\hline
		Method & Data Scenario & \text{RMSE} & Bias & \text{NMAD} \\
		\hline
		SED‑fitting & Complete & 0.243 & –0.049 & 0.030 \\
		\hline
		\multirow{4}{*}{CatBoost} & Complete (0\%)     & 0.203 & –0.001 & 0.056 \\
		& Mild (10\%)   & 0.220 & –0.001 & 0.067 \\
		& Moderate (20\%)& 0.248 & –0.001 & 0.079 \\
		& Extreme (30\%)& 0.283 & –0.001 & 0.093 \\
		\hline
	\end{tabular}
\end{table}

\subsection{Performance Comparison: ML vs. SED-Fitting}\label{sec:4.1}
We first compare the accuracy and bias of CatBoost and parametric SED‑fitting for mass, SFR, and redshift estimation under controlled missingness. The SED‑fitting reference uses complete 26‑band photometry (ideal conditions), while CatBoost is trained on only 12 bands with increasing levels of missing data (mild 10\%, moderate 20\%, extreme 30\% missing per band).

\begin{figure*}
	\centering
	\includegraphics[width=0.9\linewidth]{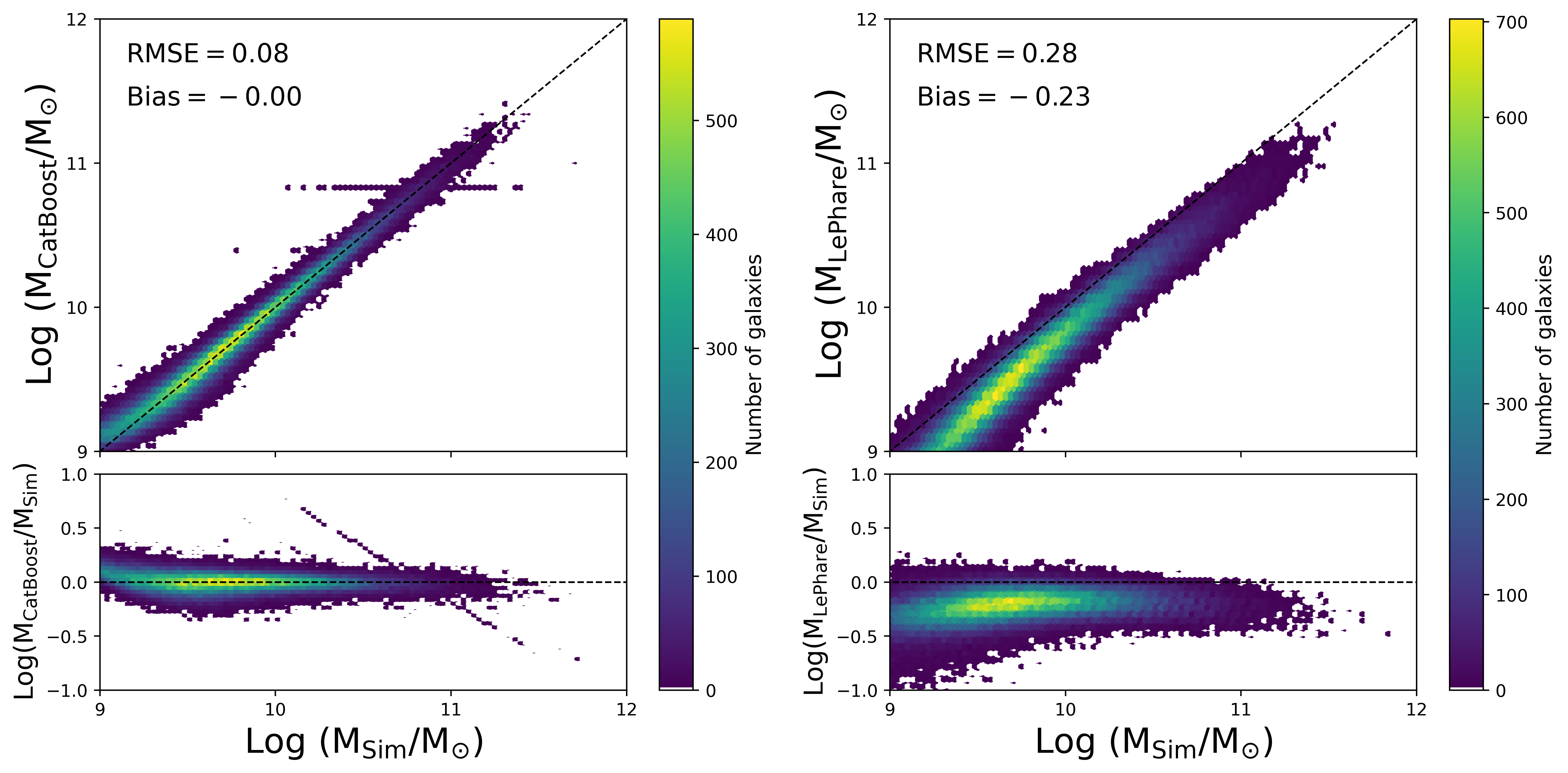}
	\caption{Comparison of mass estimates from CatBoost (trained on complete 12‑band data, left) and SED‑fitting (complete 26‑band data, right) against the true simulation masses. The displayed RMSE and bias values are in dex.}
	\label{fig:fig3}
\end{figure*}

\begin{figure*}
	\centering
	\includegraphics[width=0.95\linewidth]{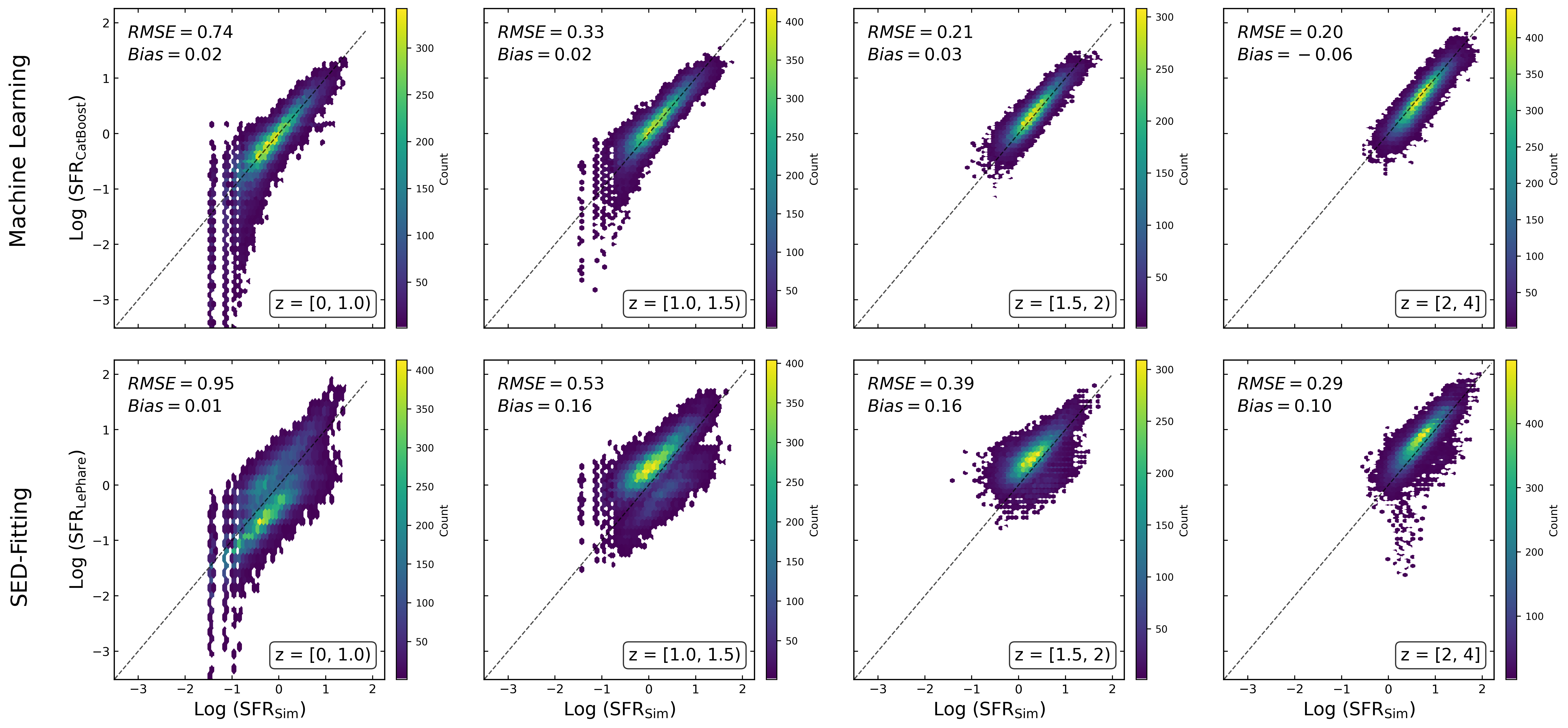}
	\caption{Comparison of logarithmic SFR estimates against true simulation values on the testing set, split into four redshift bins. Top row: CatBoost trained on complete 12‑band data. Bottom row: SED‑fitting on complete 26‑band data. The displayed RMSE and bias values are in dex.}
	\label{fig:fig4}
\end{figure*}

Figure~\ref{fig:fig3} compares mass estimates from CatBoost (complete 12‑band data, left) and SED‑fitting (complete 26‑band data, right). CatBoost achieves an RMSE of 0.079~dex and a bias of 0.000; SED‑fitting gives 0.275~dex and a bias of –0.226. When CatBoost is trained on incomplete data (Table~\ref{tab:mass_performance}), the RMSE increases gradually: 0.114~dex at mild, 0.147~dex at moderate, and 0.178~dex at extreme missingness. Bias remains near zero in all cases. At the extreme level, CatBoost’s RMSE (0.178~dex) is lower than SED‑fitting’s (0.275~dex), and its bias is smaller.

\begin{figure*}
	\centering
	\includegraphics[width=0.9\linewidth]{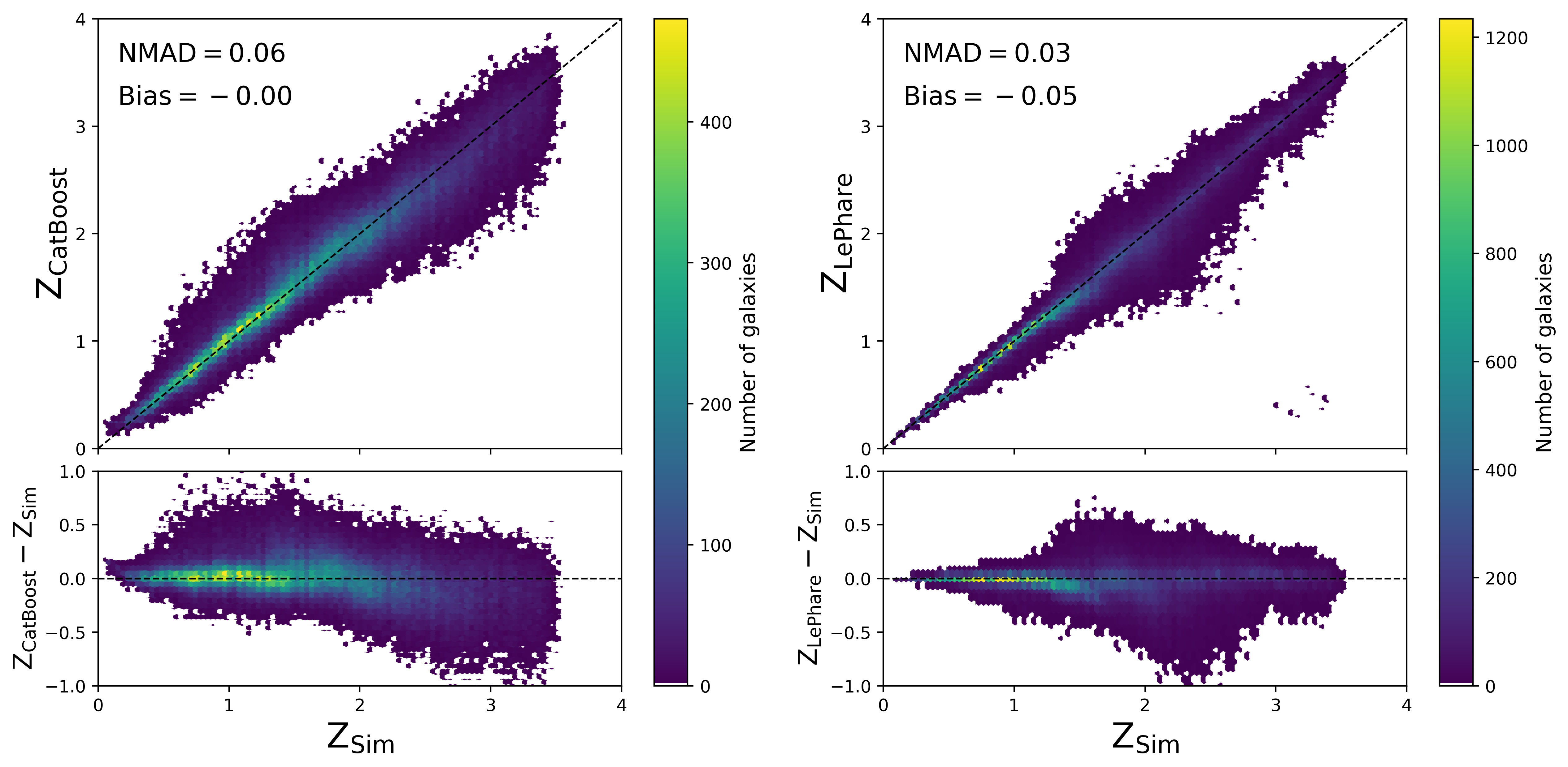}
	\caption{Comparison of redshift estimates from CatBoost (trained on complete 12‑band data) and SED‑fitting (complete 26‑band data) against the true simulation redshifts. The displayed NMAD and bias values are shown in each panel.}
	\label{fig:fig5}
\end{figure*}

Table~\ref{tab:sfr_performance} compares SFR estimates (in dex, i.e., \(\log_{10}\) SFR) from CatBoost (complete 12‑band data) and SED‑fitting (complete 26‑band data). CatBoost achieves an RMSE of 0.413~dex and a bias of –0.002; SED‑fitting gives 0.571~dex and a bias of 0.106. Figure~\ref{fig:fig4} breaks this comparison into four redshift bins. In the lowest bin (\(z\in[0,1]\)), CatBoost has an RMSE of 0.75~dex (bias +0.02) versus 0.95~dex (bias +0.01) for SED‑fitting. In \(z\in[1,1.5]\), the values are 0.33~dex (bias +0.02) vs 0.53~dex (bias +0.16); in \(z\in[1.5,2]\), 0.21~dex (bias +0.03) vs 0.39~dex (bias +0.16); and in the highest bin (\(z\in[2,4]\)), CatBoost achieves 0.20~dex (bias –0.06) while SED‑fitting reaches 0.29~dex (bias +0.10).

When CatBoost is trained on incomplete data (Table~\ref{tab:sfr_performance}), the RMSE increases gradually: 0.463~dex at mild (10\% missing), 0.497~dex at moderate (20\%), and 0.534~dex at extreme (30\%). The bias remains near zero throughout. At the extreme level, CatBoost’s RMSE (0.534~dex) is lower than SED‑fitting’s (0.571~dex), and its bias is smaller.

Table~\ref{tab:redshift_performance} compares redshift estimates from CatBoost (complete 12‑band data) and SED‑fitting (complete 26‑band data). SED‑fitting achieves a lower NMAD (0.030) compared to CatBoost (0.056), indicating better scatter performance. CatBoost has a smaller bias (–0.001) than SED‑fitting (–0.049). Figure~\ref{fig:fig5} illustrates this comparison. When CatBoost is trained on incomplete data (Table~\ref{tab:redshift_performance}), its NMAD increases to 0.067 at mild (10\% missing), 0.079 at moderate (20\%), and 0.093 at extreme (30\%). Bias remains near zero throughout.

\begin{figure*}
	\centering
	\includegraphics[width=0.91\linewidth]{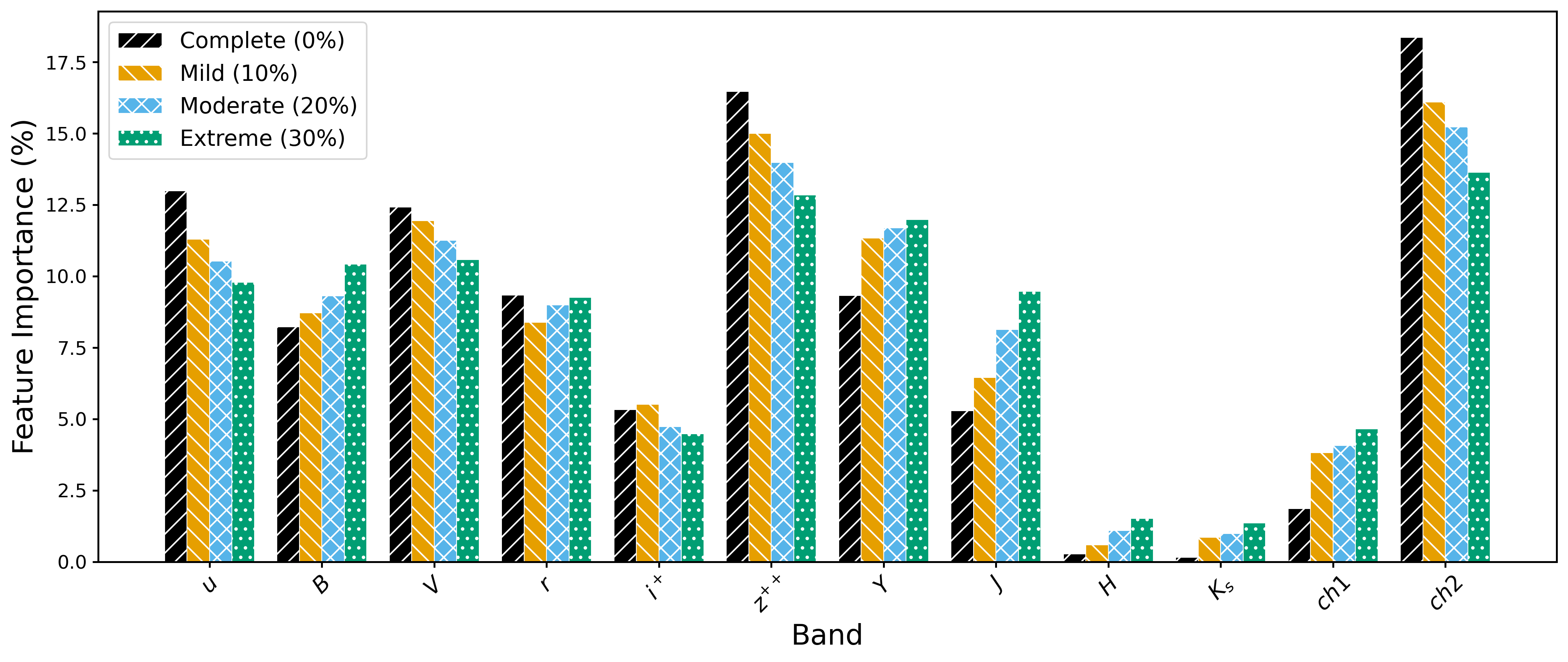}
	\caption{Feature importance of CatBoost for mass, SFR, and redshift estimation across four data conditions: complete (0\%), mild (10\% missing per band), moderate (20\% missing), and extreme (30\% missing).}
	\label{fig:fig6}
\end{figure*}

\subsection{Feature Importance Shifts with Missingness}\label{sec:4.2}
To examine how CatBoost’s internal predictions change with missing data, we analyse its feature importance. Figure~\ref{fig:fig6} shows the feature importance of each photometric band for the complete data and for the three missingness levels (mild, moderate, extreme). In the complete‑data model, the mid‑infrared ch2 band is the most important (18.4\%), followed by z$^{++}$ (16.5\%), u (13.0\%), and V (12.4\%). As missingness increases, the importance of several bands shifts. For example, the importance of B, Y, J, and ch1 rises, while that of u, V, z$^{++}$, and ch2 declines. The detailed evolution for all bands is listed in Table~\ref{tab:importance_evolution_complete}. These shifts coincide with the limited performance degradation reported in Section~\ref{sec:4.1}, indicating that the model compensates for missing information by adjusting its reliance on different bands.

\begin{table}
	\centering
	\caption{Feature importance evolution across missingness levels.}
	\label{tab:importance_evolution_complete}
	\small
	\begin{tabular}{l|cccc}
		\hline\hline
		Band & Complete & Mild & Moderate & Extreme \\
		\hline
		u     & 12.98 & 11.30 & 10.53 &  9.79 \\
		B     &  8.22 &  8.72 &  9.31 & 10.42 \\
		V     & 12.42 & 11.95 & 11.25 & 10.58 \\
		r     &  9.34 &  8.38 &  9.00 &  9.26 \\
		i$^+$ &  5.33 &  5.51 &  4.73 &  4.49 \\
		z$^{++}$& 16.47 & 15.00 & 13.98 & 12.84 \\
		Y     &  9.33 & 11.33 & 11.69 & 11.99 \\
		J     &  5.28 &  6.46 &  8.14 &  9.47 \\
		H     &  0.27 &  0.59 &  1.09 &  1.51 \\
		$\text{K}_\text{s}$    &  0.15 &  0.85 &  0.99 &  1.36 \\
		ch1   &  1.86 &  3.82 &  4.07 &  4.65 \\
		ch2   & 18.36 & 16.10 & 15.22 & 13.64 \\
		\hline
	\end{tabular}
\end{table}

\section{Discussion}\label{sec:5}
Our results show that CatBoost, when trained on data with controlled missingness (10\%, 20\%, 30\% missing per band, equal fractions across bands), shifts its feature importance towards certain bands (e.g., B, Y, J, ch1) and away from others (e.g., u, V, z$^{++}$, ch2). These shifts coincide with limited performance degradation and near‑zero bias, even at the extreme missingness level. When compared to an ideal parametric SED‑fitting reference, CatBoost yields lower RMSE and bias for mass and SFR, while for redshift it shows a trade‑off: larger scatter (NMAD) but smaller bias.

\subsection{Comparison With Parametric SED-Fitting}\label{sec:5.1}
A primary objective of this study is to evaluate whether an ML model, specifically adapted to handle incomplete photometry, can match or exceed the accuracy of a parametric SED-fitting method. Crucially, our comparison contrasts SED-fitting—which operates on complete photometry utilizing the full suite of 26 available bands—with CatBoost, which is trained under progressively severe, missingness using only the restricted 12-band set. Despite this informational disadvantage, the results show a clear advantage for the ML approach in bias for all three physical parameters, and in scatter (RMSE) for mass and SFR; for redshift the scatter (NMAD) is larger than in the SED‑fitting reference, but the bias is smaller.

For mass estimation, the ML approach shows lower random scatter and avoids the systematic bias present in the SED‑fitting results. While SED‑fitting exhibits a consistent negative bias, the CatBoost model maintains near‑zero bias even as the training data become incomplete. This is consistent with findings from \cite{asadi_mass}, who demonstrated that an ML algorithm trained on BC03 models applied to the Horizon-AGN simulation exhibited considerably less systematic negative bias than SED‑fitting. This suggests that, at least in the regime explored here, ML methods may be less affected by some of the methodological challenges that limit template-based fitting. These challenges are well-documented in the literature and include the mismatch between simplified parametric star formation histories in template libraries and the complex histories of simulated galaxies, discretization errors from sparse parameter grids, and the fundamental age-dust-metallicity degeneracy that can systematically bias recovered parameters \citep[e.g., ][]{brammer2008eazy,conroy2009propagation,Pacifici2015}. A similar tendency for SED‑fitting to underestimate mass has been noted in other simulation-based studies, such as recent work with JWST-like mock observations \citep{guzman2025synthetic}.

The improvement is most pronounced for SFR estimation. SED-fitting encounters significant challenges here, yielding considerable scatter and a large positive bias. In comparison, the ML approach substantially reduces these problems: it lowers the overall prediction error and removes the systematic bias. This pattern is consistent with findings from other work, such as the study by \cite{Davidzon2022}, which showed that SFRs inferred from self-organizing maps \citep[SOMs,][]{Kohonen1981} were significantly more consistent with independent UV–far-IR SFR measurements \citep{Barro2019} than those produced by SED-fitting. Likewise, \cite{Davidzon2019} demonstrated that SOMs applied to the Horizon-AGN simulation yield better SFR predictions than SED-fitting methods. This suggests that deriving SFRs from broad-band photometry is an area where template-based methods are more sensitive to modeling assumptions, and where more flexible, data-driven techniques may offer a clearer path to reliable estimates.

For redshift estimation, the performance gap is narrower, reflecting the inherent strength of SED-fitting in matching recognizable spectral features like the 4000 {\AA} break, which is particularly well sampled at low-redshift. The template-based method performs robustly on complete data. CatBoost trained on complete data has a slightly lower RMSE, and its performance remains stable even when substantial missingness is introduced. However, SED‑fitting achieves a lower NMAD (0.030) compared to CatBoost’s 0.056 on complete data, indicating better scatter performance.

The results suggest a practical strategy for real data: if a simulation provides a sufficiently realistic representation of the galaxy population and observational effects, training ML algorithms on the simulation and applying them to real photometry may yield more accurate and less biased estimates than running SED‑fitting directly on real data. This approach leverages the ability of ML models to learn complex correlations from simulation data, while avoiding the systematic biases that plague template‑based methods.

\subsection{Limitations and Future Work}\label{sec:5.2}
The main limitation of this study is that the SED‑fitting reference relies on parametric SFHs (exponentially declining and delayed models); other SED‑fitting codes or non‑parametric SFHs might yield different performance, and the generality of our comparison is therefore limited to the specific template set and assumptions adopted by the Horizon‑AGN catalog. Furthermore, the study depends on a single hydrodynamical simulation (Horizon-AGN); the quantitative performance of CatBoost relative to SED‑fitting may differ in other simulations (e.g., IllustrisTNG \citep{nelson2018first}, EAGLE \citep{schaye2015eagle}). Additionally, because the training and testing sets are drawn from the same simulation, the model learns from galaxies that share the same underlying physical recipes and noise properties as those in the testing set. This may make the prediction task easier than it would be for real observations, where the training sample (e.g., from a simulation) may not perfectly match the target population. Our results should therefore be interpreted as a proof of concept within a controlled simulation environment.

Additionally, our analysis is restricted to \(\text{z}<4\) and \(\log_{10} (\text{M}) > 9\); the results may not generalise to higher‑redshift or lower‑mass galaxies. Our missingness injection assumes independence across bands and a fixed transition width \(\alpha=1.5\), whereas real surveys exhibit correlated non‑detections and wavelength‑dependent completeness. This simplification was deliberate to provide a controlled, interpretable baseline. The quantitative performance metrics (e.g., exact RMSE values) may not directly generalise to surveys with strong correlated missingness, but the main qualitative results (graceful degradation, near‑zero bias, the main qualitative trends) are likely robust. Future work should test CatBoost under more realistic correlated missingness models and on real data.

\section{Conclusion}\label{sec:6}
This study provides a deliberately asymmetric stress test of CatBoost against parametric SED‑fitting for galaxy mass, SFR, and redshift estimation under controlled photometric incompleteness (mild 10\%, moderate 20\%, extreme 30\% missing per band). Using a mock Horizon-AGN catalog and controlled missingness injection, we show that CatBoost’s native missing‑value handling leads to limited performance degradation and near‑zero bias even at the extreme missingness level. When compared to an ideal parametric SED‑fitting reference (complete 26‑band data), CatBoost trained on only 12 bands with 30\% missing values achieves lower errors for mass and SFR and eliminates systematic biases. The principal numerical findings are:

\begin{itemize}
	\item \textit{Limited performance degradation under severe missingness:} As missingness increases from 0\% to 30\%, CatBoost maintains good accuracy. Mass RMSE rises from 0.08 to 0.18~dex, SFR RMSE from 0.41 to 0.53~dex, and redshift RMSE from 0.20 to 0.28. Bias remains near zero throughout.
	
	\item \textit{Comparison with parametric SED‑fitting:} For mass and SFR, CatBoost outperforms parametric SED‑fitting in RMSE and bias across all missingness levels. For example, at the extreme level, CatBoost’s mass RMSE is 0.18 vs 0.28~dex, and SFR RMSE is 0.53 vs 0.57~dex. For redshift, parametric SED‑fitting achieves lower NMAD (e.g., 0.030 vs 0.093 at extreme) while CatBoost has smaller bias (–0.001 vs –0.049).
\end{itemize}

These results demonstrate that CatBoost, with its native handling of missing data, can provide more accurate and less biased estimates than a parametric SED‑fitting method for galaxy physical properties under controlled missingness, at least for the simulated conditions explored here. This suggests that CatBoost may offer practical advantages for extracting galaxy properties from real photometric surveys.

\section*{Data Availability}
The Horizon-AGN simulation data are publicly accessible at \url{https://www.horizon-simulation.org/data.html}.

\bibliography{ref}

\end{document}